\documentclass[final,number,sort&compress,11pt]{elsarticle}
\usepackage{amsmath,amssymb,array,bm,comment,epsfig,epstopdf,graphicx,lineno,multirow,setspace,siunitx,url,xcolor}
\usepackage[letterpaper,top=1.25in,bottom=1.25in,left=1in,right=1in]{geometry}
\usepackage[version=3]{mhchem}

\newcommand{\eqphase}[1]{{\ensuremath\underline{#1}}}
\DeclareSIUnit\inch{in}

\begin{document}

\title{Application of Finite Element, Phase-field, and CALPHAD-based Methods to Additive Manufacturing of \ce{Ni}-based Superalloys}

\author[mse]{Trevor~Keller\corref{cor}}
\ead{trevor.keller@nist.gov}
\author[mse]{Greta~Lindwall}
\author[mse]{Supriyo~Ghosh}
\author[mse,theiss]{Li~Ma}
\author[eng]{Brandon~M.~Lane}
\author[mms]{Fan~Zhang}
\author[mse]{Ursula~R.~Kattner}
\author[mse]{Eric~A.~Lass}
\author[eng]{Jarred~C.~Heigel}
\author[mse]{Yaakov~Idell}
\author[mse]{Maureen~E.~Williams}
\author[mms]{Andrew~J.~Allen}
\author[mse]{Jonathan~E.~Guyer}
\author[mse]{Lyle~E.~Levine}

\cortext[cor]{Corresponding author.}

\address[mse]{Materials~Science~and~Engineering~Division,
              Material~Measurement~Laboratory,
              National~Institute~of~Standards~and~Technology,
              100 Bureau Drive,
              Gaithersburg, Maryland 20899, U.~S.~A.}
\address[theiss]{Theiss Research,
                 7411 Eads Avenue,
                 La Jolla, CA 92037}
\address[eng]{Intelligent~Systems~Division,
              Engineering~Laboratory,\\
              National~Institute~of~Standards~and~Technology,
              100 Bureau Drive,
              Gaithersburg, Maryland 20899, U.~S.~A.}
\address[mms]{Materials Measurement Science Division,
              Material~Measurement~Laboratory,
              National~Institute~of~Standards~and~Technology,
              100 Bureau Drive,
              Gaithersburg, Maryland 20899, U.~S.~A.}

\date{May~1, 2017}

\begin{abstract}
	Numerical simulations are used in this work to investigate aspects of
	microstructure and microsegregation during rapid solidification of a
	\ce{Ni}-based superalloy in a laser powder bed fusion additive manufacturing process.
	Thermal modeling by finite element analysis simulates the laser melt pool,
	with surface temperatures in agreement with \emph{in situ} thermographic
	measurements on Inconel~625. Geometric and thermal features of the
	simulated melt pools are extracted and used in subsequent mesoscale simulations.
	Solidification in the melt pool is simulated on two length scales.
	For the multicomponent alloy Inconel 625, microsegregation between dendrite
	arms is calculated using the Scheil-Gulliver solidification model and DICTRA software.\footnote{
		Certain commercial entities, equipment, or materials may be identified in this document in order to describe
		an experimental procedure or concept adequately. Such identification is not intended to imply recommendation
		or endorsement by the National~Institute~of~Standards~and~Technology (NIST), nor is it intended to imply that the
		entities, materials, or equipment are necessarily the best available for the purpose.}
	Phase-field simulations, using \ce{Ni}--\ce{Nb} as a binary analogue to Inconel~625,
	produced microstructures with primary cellular/dendritic arm spacings in agreement
	with measured spacings in experimentally observed microstructures and a lesser extent
	of microsegregation than predicted by DICTRA simulations.
	The composition profiles are used to compare thermodynamic driving forces for nucleation
	against experimentally observed precipitates identified by electron and X-ray
	diffraction analyses. Our analysis lists the precipitates that may form from FCC phase of
	enriched interdendritic compositions and compares these against experimentally observed
	phases from \SI{1}{\hour} heat treatments at two temperatures:
	stress relief at  \SI{1143}{\kelvin} (\SI{870}{\degreeCelsius}) or
	homogenization at \SI{1423}{\kelvin} (\SI{1150}{\degreeCelsius}).
\end{abstract}


\maketitle

\section{Introduction}
	Inconel~625 (IN625) is a \ce{Ni}-based superalloy used for turbine parts that is
	strengthened by substitutional alloying elements such as \ce{Cr}, \ce{Mo}, and \ce{Nb}.
	Laser powder bed fusion (L-PBF), an additive manufacturing technique,
	presents opportunities to reduce the cost of making IN625 parts
	with appropriate geometries and internal cooling channels for high-temperature applications.
	Heat treatment is often necessary following additive manufacturing to relieve residual stress
	\cite{Denlinger2015,Tremsin2016}, and to homogenize the microstructure \cite{Xu2013,Mostafaei2016}.
	Recent work shows that common heat treatments promote precipitation of secondary phases \cite{Zhang2017},
	which degrade mechanical properties (such as indentation hardness) in IN625 \cite{Suave2014}.
	Heat treatment schedules for wrought IN625 were designed to avoid these same precipitates \cite{Floreen1994};
	however, there are substantial microstructural differences between wrought and L-PBF material \cite{Idell2016},
	with significant microsegregation of as-solidified material of particular interest here.
	IN625 processed by welding, casting, or directional solidification exhibits less homogeneity than wrought material,
	and typically contains \ce{NbC} and Laves phase precipitates \cite{DuPont1996,Formenti2005}.
	Finding suitable stress-relieving and homogenizing heat treatments without sacrificing strength
	is an iterative process, but numerical modeling can help narrow the search.

	3D finite element modeling has been crucial to understanding the L-PBF process.
	Models describing single laser tracks across the powder bed surface are routinely used
	to simulate heat dissipation through the solid substrate \cite{Song2012,Yin2012}.
	Multi-track models allow for coupling residual stress evolution to these thermal profiles \cite{Li2010,Hussein2013,Hodge2014}.
	Recent models improve the thermal modeling using coupled Calculation of Phase Diagrams (CALPHAD) methods for
	accurate prediction of the melt pool boundary and solidification microstructure \cite{Smith2016}.
	Such advances in modeling the L-PBF system configuration and relevant materials properties
	are critical to accurate finite element analysis (FEA) predictions of the real process under conditions of interest.

	Phase-field methods are commonly used to simulate microstructural features,
	between atomistic and continuum length scales.
	A scalar-valued order parameter defined throughout a spatial domain is
	used in these methods to label the presence or absence of some phase,
	hence ``phase field.''
	The best available phase-field models for solidification were performed in 2D for binary alloys
	\cite{Karma1998,Echebarria2004}.
	The simplified geometry and approximation of multicomponent alloys as binaries
	allows for efficient computation, and the models produce quantitatively correct
	mass redistribution across the solidifying interface.
	The model has been applied successfully to tungsten arc welding of \ce{Al}--\ce{Cu} \cite{Farzadi2008},
	laser powder forming of \ce{Ti}--\ce{Nb} \cite{Fallah2012} and \ce{Ni}--\ce{Nb} \cite{Nie2014},
	and electron beam additive manufacturing of \ce{Ti}--\SI{6}{\percent} \ce{Al}--\SI{4}{\percent} \ce{V}
	by combining \ce{Al} with \ce{V} as a virtual element with mass fraction \SI{10}{\percent} \cite{Gong2015}.
	
	A third approach to studying microsegregation during solidification uses DICTRA software \cite{Andersson2002},
	which implements a one-dimensional model for diffusion-controlled phase transformation.
	Despite its simplicity, this approach has the advantage of allowing for 
	simulation of microsegregation and back-diffusion in multicomponent materials by combining
	CALPHAD thermodynamic and kinetic materials descriptions.
	DICTRA is routinely applied to multicomponent alloys,
	including \ce{Ni}-based superalloys \cite{Jablonski2009}.

	These modeling techniques---FEA, phase-field, and CALPHAD-based---are already in use
	studying various aspects of L-PBF, separately and increasingly in cooperation,
	for a variety of alloy systems.
	It is our goal to integrate all three models together to achieve high fidelity simulations
	of dendritic solidification in L-PBF IN625, with direct comparison against experimental results.
	This effort will improve the fundamental understanding of solidification in this system
	and produce input data for modeling solid state transformations in the future.

\section{Numerical methods}
	\subsection{Finite element thermal model\label{sec:meth_thermal}}
\begin{figure*}[tp]\centering
		    \includegraphics[width=\textwidth]{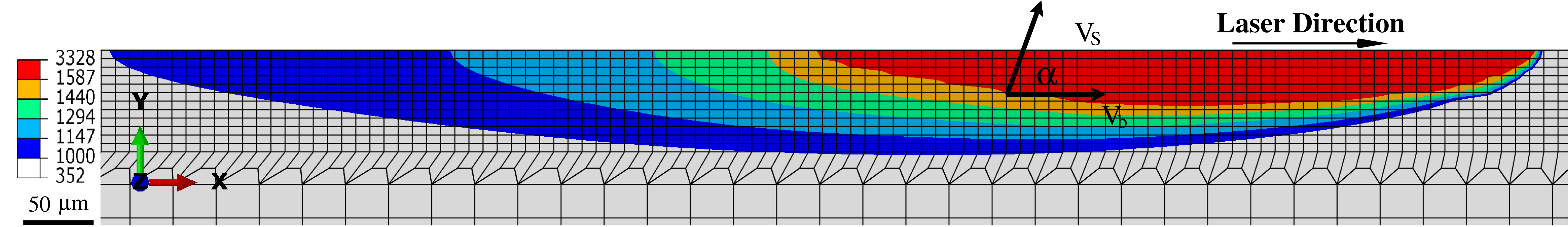}
		    \caption{2D section through melt pool centerline in 3D FEA simulation of IN625 L-PBF using a
		             single layer of powder on a bulk substrate showing thermal profile into the substrate.
		             Dividing line between red and orange contours represents the solid-liquid interface,
		             from which transformation angle $\alpha$ and solidification speed
		             $V_s = V_b\cos\alpha$ are calculated as shown. Indicated temperatures are measured in \si{\kelvin}.}
		    \label{fig:fea_meltpool}
		\end{figure*}

		Using the commercial FEA code ABAQUS \cite{Abaqus2013},	a non-linear, transient, thermal
		model was designed and executed to obtain the global temperature history
		generated during laser irradiation of one layer of powder covering a solid substrate.
		The simulated powder layer thickness is \SI{36}{\micro\meter},
		combining the nominal \SI{20}{\micro\meter} layer height with
		\SI{16}{\micro\meter} of underlayer densification.
		This matches the steady-state powder layer thickness observed in
		corresponding experimental builds.
		Both single-track and multiple-track laser scans across the metal powder layer were modeled,
		in which one ``track'' is a linear path to be followed by the laser separated by
		\SI{100}{\micro\meter} from adjacent laser pathways traversed in anti-parallel directions.
		The width of the molten pool in the transverse direction is \SI{140}{\micro\meter},
		so material at the midpoint melts on adjacent scans.
		To reduce computation time, the elements that interact with the laser beam are
		finely meshed within the diameter of the laser, and a coarse mesh was used for
		the surrounding loose powder and substrate, visible in Figure~\ref{fig:fea_meltpool}.
		Ma \emph{et al.} \cite{Ma2015} described this model previously,
		with detailed discussion of appropriate parameter values.

		Heat transfer in the L-PBF process was modeled using the energy balance equation with
		Fourier's law of heat conduction and internal sources of heat \cite{Bird2007},
		\begin{equation}\label{eqn:fourier}
			\frac{\partial\left(\rho c_p T\right)}{\partial t} = \nabla\cdot\left(\kappa\nabla T\right) + Q,
		\end{equation}
		in which the evolution of temperature $T$ with time $t$ depends on
		material density $\rho$,
		thermal conductivity $\kappa$,
		specific heat capacity $c_p$ which depends on latent heat,
		and internal heat $Q$ which depends on radiative loss.
		The temperature-dependent bulk material density and specific heat were calculated from a
		Scheil-Gulliver simulation for the nominal IN625 composition and using the TCNI8 thermodynamic
		database \cite{TCNI8} within the Thermo-Calc software \cite{Andersson2002}.
		The initial condition assumed a uniform temperature of \SI{353}{\kelvin} throughout the specimen at time $t = 0$.
		Adiabatic conditions were applied to all boundaries except the top surface,
		on which the boundary condition is
		\begin{equation}\label{eqn:thermalbc}
			\left(-\kappa\nabla T\right)\cdot\hat{n} = q_s + h\left(T-T_e\right) + \varepsilon\sigma\left(T^4 - T_e^4\right).
		\end{equation}
		The three terms on the right-hand side represent heat input from the laser,
		heat convection due to flowing process gas, and radiation.
		Equation~\ref{eqn:thermalbc} depends on the surface normal $\hat{n}$, laser input heat $q_s$,
		convective heat transfer coefficient $h$, thermal radiation coefficient $\varepsilon$,
		the Stefan-Boltzmann constant $\sigma$, and ambient temperature $T_e$.
		The laser input was modeled after the single-mode continuous wave \ce{Yb} fiber laser ($\lambda = \SI{1070}{\nano\meter}$)
		used in our experimental L-PBF system (described in Sec.~\ref{sec:expt}).
		Interaction between the laser and material is modeled using a Gaussian expression for surface heat flux \cite{Roberts2009},
		\begin{equation}\label{eqn:gauss}
			q_s = \frac{2AP}{\pi r_b^2}\exp\left(\frac{-2r^2}{r_b^2}\right),
		\end{equation}
		with power $P = \SI{195}{\watt}$, powder bed absorption coefficient $A = \num{0.50}$,
		laser beam radius $r_b = \SI{50}{\micro\meter}$, and radial distance to the beam centerline
		$r$ measured in \si{\micro\meter}.
		The simulated laser scanning speed was \SI{0.8}{\meter\per\second}.
			
		Each element stores its temperature and a Boolean variable indicating whether it
		has ever exceeded the liquidus temperature, $T_\ell$.
		Elements in the powder layer are initialized with this variable set to ``false,''
		indicating a powder state.
		Upon melting, the variable switches to ``true,'' indicating a bulk state.
		Substrate elements are initialized with this variable set to ``true.''
		There is no mechanism for switching this melt-state variable from ``true'' to ``false:''
		the fused material can never revert to powder.
		The materials parameters $\rho$ and $\kappa$ for each element depend on both variables, $T$ and melt-state.
		In the powder state, $\kappa$ depends on the packing fraction, particle size distribution,
		particle morphology, and thermal conductivity of the bulk material and process gas \cite{Rombouts2005,Tanaka2012}.
		In the simulations described in this work, $\kappa$ was specified in the range from
		\SI{1.0}{\watt\per\meter\per\kelvin} to \SI{3.0}{\watt\per\meter\per\kelvin} after \cite{Childs2005}.
		As $T$ rises during the first melting event, $\rho$ and $\kappa$ linearly increase
		from their powder to bulk values when $T$ is above the solidus temperature $T_s$ and below the
		liquidus temperature $T_\ell$ given by the IN625 phase diagram \cite{TCNI8,Andersson2002}.
		If $T$ exceeds $T_s$, then the melt-state variable switches to ``true;'' 
		$\rho$ and $\kappa$ are thereafter functions of $T$, only.
		Note that the bulk state variable does not differentiate between solid and liquid phases:
		the FEA model uses $T$ to choose phase-dependent materials properties appropriately.
		Latent heat effects are captured in the evaluation of $c_p$ \cite{Abaqus2013}.

	\subsection{CALPHAD-based solidification models\label{sec:meth_thermo}}
		To estimate the extent of microsegregation during solidification of a material, the
		Scheil-Gulliver model \cite{Gulliver1909,Scheil1942} is often applied.
		The model can be used for multicomponent materials, provided that a thermodynamic
		description for the multicomponent alloy is available. The model assumes perfect mixing in the
		liquid and no diffusion in the solid phase.
		These conditions are not found in nature, so the result is a theoretical limit:
		less segregation is expected during real solidification processes than the Scheil-Gulliver model predicts,
		since finite diffusion in both the liquid and solid phases will contribute to mass redistribution.

		Microsegregation predictions may also be obtained using DICTRA software \cite{Andersson2002}
		that include the effect of diffusion in liquid and solid during solidification.
		Local equilibrium is assumed at the interface between phases, 
		in this case liquid and $\gamma$, and flux-balance is maintained for each element.
		Solutions to the diffusion equation	and the assumption of local equilibrium at the phase interface are used
		to determine the tie-line that satisfies flux-balance.
		Because the simulations are 1D, the effects of dendrite tip diffusion are not included.

		For the IN625 segregation simulation, we used the commercial thermodynamic database TCNI8 \cite{TCNI8} and the
		NIST \ce{Ni} Superalloy mobility database \cite{Campbell2002}.
		The DICTRA simulation domain was \SI{150}{\nano\meter}, which is half of the secondary dendrite arm spacing
		measured from experimental microstructures (e.g., Figure~\ref{fig:experiment}).
		Temperature was specified as spatially uniform, but time-dependent,
		with values taken directly from the FEA thermal model.
		To reduce the computational complexity, metals contributing mass fractions below \SI{0.5}{\percent} were excluded,
		producing a	simplified system composition
		\ce{Ni}--\SI{0.1}{\percent} \ce{C}--\SI{20.11}{\percent} \ce{Cr}--\SI{0.72}{\percent} \ce{Fe}--\SI{8.83}{\percent} \ce{Mo}--\SI{3.75}{\percent} \ce{Nb}.

	\subsection{Phase-field solidification model}
		We used a quantitative phase-field model in 2D to study the cellular/dendritic nature of the solidification
		of the melt pool \cite{Echebarria2004},	with a simplified representation of IN625 as a \ce{Ni}--\ce{Nb}
		binary system with only FCC $\gamma$ phase and liquid phase.
		The bulk composition has a mass fraction of \SI{4}{\percent} \ce{Nb}.
		The scalar phase parameter $\phi$ indicates whether a point in the 2D field is liquid $(\phi=-1)$,
		solid $(\phi=1)$, or within the solid-liquid interface $(-1<\phi<1)$.
		The phase-field is not conserved, and evolves in time $t$ and space $(x,y)$ as 
		\begin{align}\label{eqn:phi}
		\frac{\partial \phi}{\partial t} &= \frac{1}{\tau_0{a(\hat{n})}^2}\bigg(W_{0}^2\nabla \left[{a(\hat{n})}^2 \nabla\phi\right] + \phi - \phi^3\\\nonumber
		                                 &- \frac{\partial}{\partial x} \left[ a(\hat{n}) \frac{\partial a(\hat{n})}{\partial\hat{n}} \frac{\partial\phi}{\partial y}\right]
		                                  + \frac{\partial}{\partial y} \left[ a(\hat{n}) \frac{\partial a(\hat{n})}{\partial\hat{n}} \frac{\partial\phi}{\partial x}\right]\\\nonumber
		                                 &- \frac{\lambda}{1-k} (1-\phi^2)^2 \left[\exp\left(u\right) - 1 + \frac{T- T_0}{m_{\ell} c_0/k}\right]\bigg).
		\end{align}
		The dimensionless interfacial energy $a(\hat{n}) = 1 + \epsilon_4 \cos (4\theta)$, with four-fold anisotropy of magnitude $\epsilon_4$,
		interface normal vector $\hat{n} = -\frac{\nabla\phi}{|\nabla \phi|}$, and orientation angle
		$\theta = \arctan\left(\frac{\partial\phi}{\partial y}/\frac{\partial\phi}{\partial x}\right)$.
		The non-dimensional deviation of chemical potential, $u = \ln\left(\frac{2ck/c_0}{1+k-(1-k)\phi}\right)$,
		is defined with respect to the equilibrium chemical potential at a reference temperature $T_0$ and system composition $c_0$.
		The frozen temperature approximation is applied such that a linear temperature profile with constant gradient $G$
		translates along the growth axis $(x)$ with constant velocity $V_s$ \cite{Smith1955,Mullins1964,Langer1980,Merchant1990,Huntley1993}:
		$T(x,t) = T_0 + G(x - V_s t)$.
		Interface thickness $W_0$ and relaxation time constant $\tau_0$ are related through the 
		capillary length, $d_0 = a_1 W_0/\lambda$.
		Asymptotic analysis, performed by enforcing local equilibrium at the interface as its width vanishes,
		also links these quantities through a dimensionless coupling parameter $\lambda$
		and diffusion constant in the liquid $D_{\ell}$ \cite{Karma1996,Echebarria2004}:
		$\tau_0 = a_2\lambda W_{0}^{2}/D_{\ell}$.
		The fitting parameters $a_1 = 0.8839$ and $a_2 = 0.6267$ depend on the forms
		of the free energy functional and free energy density, respectively \cite{Karma1998}.
		$W_0$ is therefore the only free parameter, chosen to be \SI{10}{\nano\meter}.

		Composition is modeled with a conserved field $c$, and evolves as
		\begin{align}\label{eqn:conc}
		\frac{\partial c}{\partial t} &= \nabla \cdot \left( \frac{1-\phi}{2} D_\ell \left[ 1+k-(1-k)\phi\right] \frac{c_0}{k} \nabla \exp\left(u\right)\right.\\\nonumber
		                              &+ \left.\frac{W_0}{2\sqrt{2}} (1-k) \frac{c_0}{k} \exp\left(u\right) \frac{\partial \phi}{\partial t} \hat{n}\right).
		\end{align}
		This expression neglects the effect of thermal gradients on diffusion, or the Soret effect \cite{Groot1984a},
		which contributes to macrosegregation under low solidification velocities \cite{Zheng1998},
		but not microsegregation during rapid solidification.
		The phase diagram of the \ce{Ni}--\ce{Nb} system exhibits a practically linear liquidus
		with constant slope $m_{\ell}=\SI{-10.5}{\kelvin\per\percent}$, measured with respect to mass percentage \ce{Nb},
		and constant partition coefficient $k=0.48$ in this dilute region.
		Equations~\ref{eqn:phi}~and~\ref{eqn:conc} were solved on a uniform rectilinear grid using a finite volume
		method and an explicit time marching scheme with zero-flux boundary conditions.
		Model parameters for a dilute solution of \ce{Nb} in \ce{Ni} were used directly from Nie~\emph{et al.}
		\cite{Nie2014}, summarized in their Table~1.
		Further analysis of solidification microstructures in dilute \ce{Ni}--\ce{Nb} alloys using this model
		are reported elsewhere \cite{Ghosh2017}.

\section{Experimental methods and results\label{sec:expt}}
		Test cubes of IN625 were additively manufactured by the NIST Engineering Laboratory
		using an EOSINT M270 (EOS GmbH, Krailling, Germany);
		for L-PBF system details, the interested reader may refer to \cite{Lane2016}.
		The EOS NickelAlloy~IN625 powder (EOS GmbH, Krailling, Germany)
		was supplied with compositions listed in Table~\ref{tab:AM625comp}, as measured by
		inductively coupled plasma atomic emission spectroscopy (ICP) and
		flame atomic absorption spectroscopy (FAAS),
		or by X-ray fluorescence spectroscopy (XRF, conforming to \cite{ASTMxrf}).
		All measured values are within the standard ranges for IN625 \cite{ASTM625}.
		For calibration of the FEA thermal model, multiple-track laser scans were made on
		\SI{1.27}{\centi\meter} (\SI{0.5}{\inch}) thick	solutionized IN625 plate
		(High Performance Alloys, Inc. Windfall, IN, USA).
		For both IN625 media, the \ce{Yb} fiber laser power was \SI{195}{\watt} and scan speed was \SI{0.8}{\meter\per\second}.

		\begin{table}\centering
			\caption{Allowable and measured mass fractions of constituent elements for IN625 L-PBF feedstock used in this work \cite{ASTM625},
			         as determined by ICP, FAAS (indicated by an asterisk, $^*$), or XRF.}
			\label{tab:AM625comp}
			
			\begin{tabular}{ccll}\hline
				 Element & Standard Range                              & Supplied (ICP)               & Supplied (XRF)            \\\hline
				 \ce{Ni} & \SI{58.0}{\percent}  minimum                & balance                      & balance                     \\
				 \ce{Cr} & \SI{20.0}{\percent}  to \SI{23.0}{\percent} & \SI{20.7 }{\percent}         & \SI{21.1 }{\percent}        \\
				 \ce{Mo} &  \SI{8.0}{\percent}  to \SI{10.0}{\percent} & \SI{ 8.83}{\percent}         & \SI{ 6.69}{\percent}        \\
				 \ce{Nb} &  \SI{3.15}{\percent} to \SI{4.15}{\percent} & \SI{ 3.75}{\percent}         & \SI{ 3.06}{\percent}        \\
				 \ce{Fe} &  \SI{5.0}{\percent}  maximum                & \SI{ 0.72}{\percent}         & \SI{ 0.01}{\percent} maximum\\
				 \ce{Ti} &  \SI{0.4}{\percent}  maximum                & \SI{ 0.35}{\percent}         & \SI{ 0.01}{\percent} maximum\\
				 \ce{Al} &  \SI{0.4}{\percent}  maximum                & \SI{ 0.28}{\percent}         & \SI{ 0.17}{\percent}        \\
				 \ce{Co} &  \SI{1.0}{\percent}  maximum                & \SI{ 0.18}{\percent}         & \SI{ 1.01}{\percent} maximum\\
				 \ce{Si} &  \SI{0.5}{\percent}  maximum                & \SI{ 0.13}{\percent}         & \SI{ 0.13}{\percent}        \\
				 \ce{Mn} &  \SI{0.5}{\percent}  maximum                & \SI{ 0.03}{\percent}         & \SI{ 0.02}{\percent}        \\
			  \ce{C}$^*$ &  \SI{0.1}{\percent}  maximum                & \SI{ 0.01}{\percent}         &                             \\
			  \ce{P}$^*$ & \SI{0.015}{\percent} maximum                & \SI{ 0.01}{\percent} maximum & \SI{ 0.01}{\percent} maximum\\
			  \ce{S}$^*$ & \SI{0.015}{\percent} maximum                & \SI{0.002}{\percent}         &                             \\
				\hline
			\end{tabular}
		\end{table}

		Scanning electron microscopy (SEM) specimens were polished to a \SI{1}{\micro\meter} surface
        finish using standard metallographic preparation techniques,
        then etched for \SI{30}{\second} in aqua regia.

\begin{figure}[tp]\centering
		    \includegraphics[width=\textwidth]{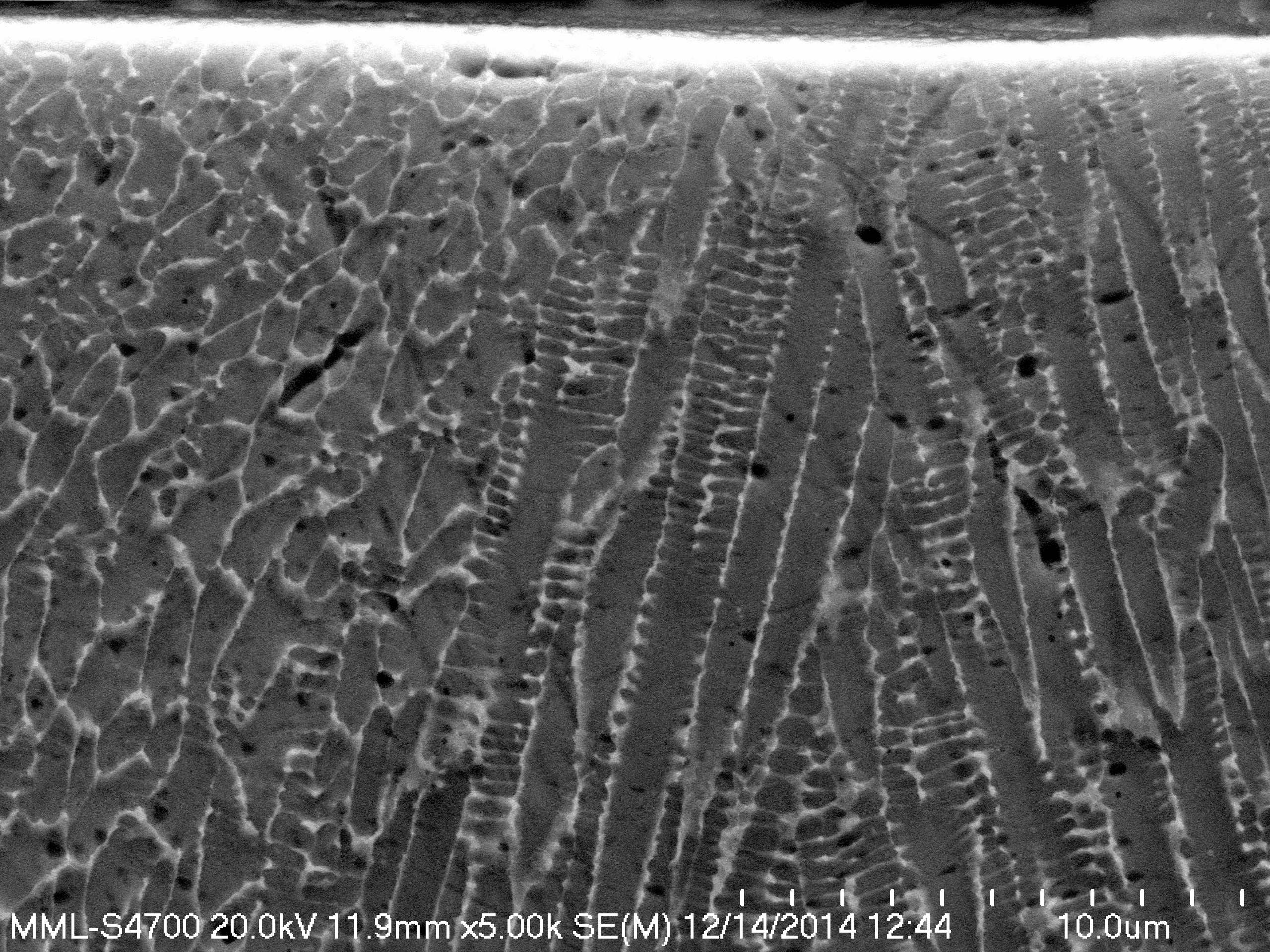}
		    \caption{Representative cross section through as-built L-PBF IN625 specimen,
		             etched with aqua regia, showing cellular/dendritic microstructure.
		             Hitachi S-4700 secondary electron image with \SI{20}{\kilo\volt} accelerating potential,
		             \SI{11.9}{\milli\meter} working distance.
		             This image has been manipulated to increase its contrast.}
		    \label{fig:experiment}
		\end{figure}

\begin{figure*}[tp]\centering
		    \includegraphics[width=\textwidth]{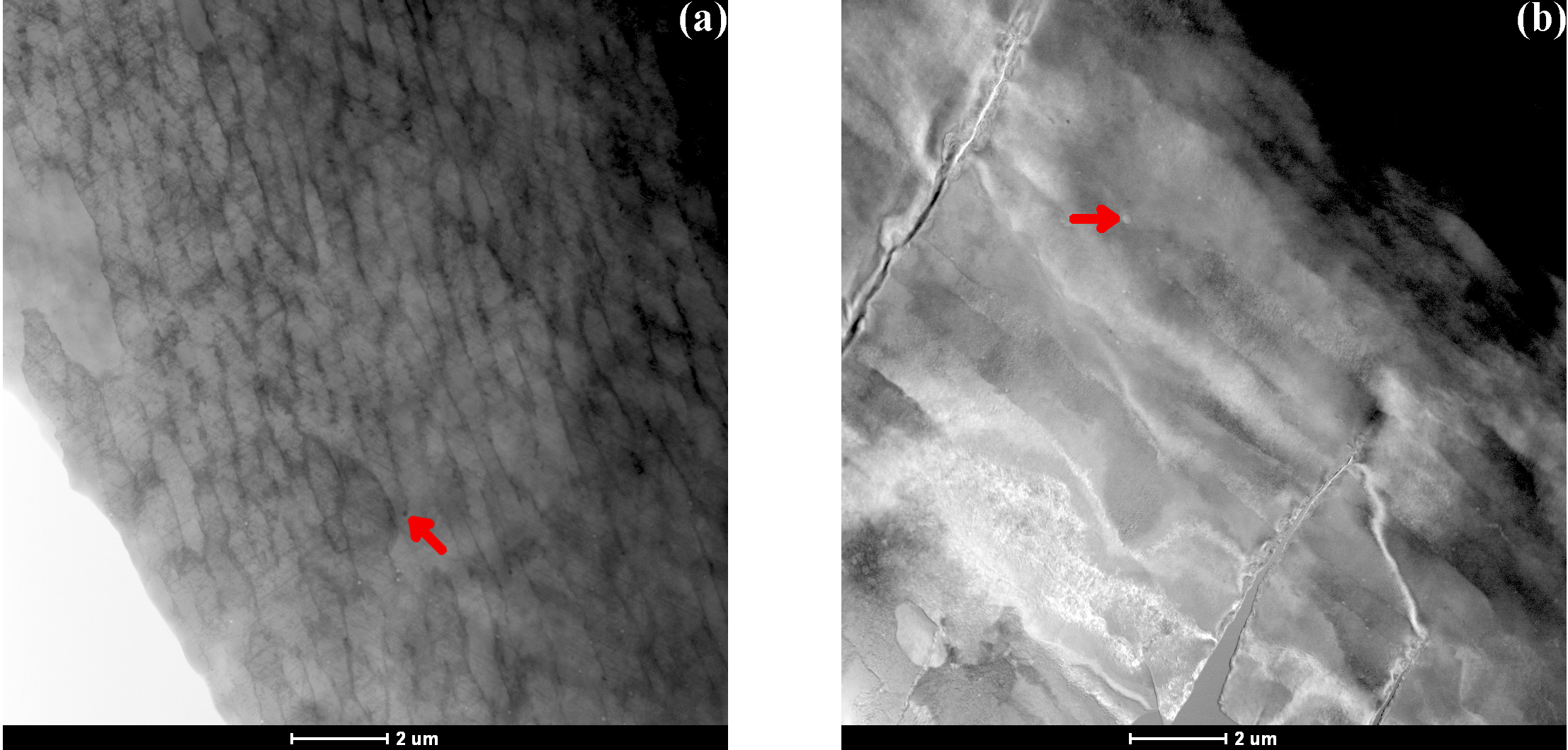}
		    \caption{Representative micrographs showing cellular/dendritic microstructure in L-PBF IN625
		             (a) as-built and
		             (b) following \SI{1}{\hour} anneal at \SI{1143}{\kelvin}.
		             Circular spots between \SI{50}{\nano\meter} to \SI{200}{\nano\meter} in diameter represent precipitates,
		             with arrows indicating an exemplar in each image.
		             Precipitates in annealed material were indexed as carbides,
		             but those in as-built material could not be uniquely identified.
		             FEI Titan 80-300 bright field STEM images with \SI{300}{\kilo\volt} accelerating potential
		             and zone axes parallel to $[1\,1\,0]$.
		             These images have been manipulated to increase contrast.}
		    \label{fig:temcarbides}
		\end{figure*}

		Specimens for transmission electron microscopy (TEM) were prepared using electrical discharge
		machining to cut disks \SI{3}{\milli\meter} in diameter,
		which were then electropolished until electron-transparent with a \SI{6}{\percent} perchloric acid,
		\SI{60}{\percent} methanol, and \SI{34}{\percent} butanol solution.
		Figure~\ref{fig:experiment} provides a representative secondary electron micrograph from the as-built L-PBF IN625 specimen:
		the primary dendrite arms are spaced \SI{1}{\micro\meter} and the secondary arms are spaced \SI{300}{\nano\meter}.
		Regions without secondary arms are also present:
		the wide range of localized solidification conditions produced by L-PBF can produce regions of cellular growth,
		so we describe the experimental microstructures as ``cellular/dendritic.''
		Figure~\ref{fig:temcarbides} provides representative bright field scanning transmission electron micrographs (STEM)
		showing precipitates as light and dark circular spots, \SI{50}{\nano\meter} to \SI{200}{\nano\meter} in diameter.
		Precipitates are found near cell/dendrite boundaries in as-built material,
		but are not apparently localized to microstructural features after stress relief at \SI{1143}{\kelvin} for \SI{1}{\hour}.
		Electron diffraction patterns confirm the presence of \ce{MC}, \ce{M6C}, and \ce{M23C6} carbides in stress-relieved material,
		whereas diffraction patterns of precipitates in as-built material could not be indexed due to residual stress and high dislocation densities.

		X-ray diffraction experiments were performed at the  ultra-small angle X-ray scattering (USAXS)
		facility at the Advanced~Photon~Source at Argonne National Laboratory \cite{Ilavsky2009,Ilavsky2013},
		with specimen preparation and measurement conditions reported previously \cite{Zhang2017}.
		The relative distance and tilt between the specimen and detector are calibrated using \ce{LaB6} powder.
		The sample was heated to and held at \SI{1143}{\kelvin} for \SI{1}{\hour}, 
		then cooled to \SI{303}{\kelvin},
		corresponding to the manufacturer's recommended stress-relieving heat treatment for IN625 \cite{EOS625}.
		The sample was then heated to and held at \SI{1423}{\kelvin} for \SI{1}{\hour},
		then cooled back to \SI{303}{\kelvin}, a treatment suggested by previous work \cite{Zhang2017}.
		The heating rates were \SI{100}{\kelvin\per\minute}, the cooling rates were \SI{200}{\kelvin\per\minute},
		and the temperature stability was \SI{\pm 1}{\kelvin}.

		\begin{figure}[tp]\centering
			\includegraphics[width=\textwidth]{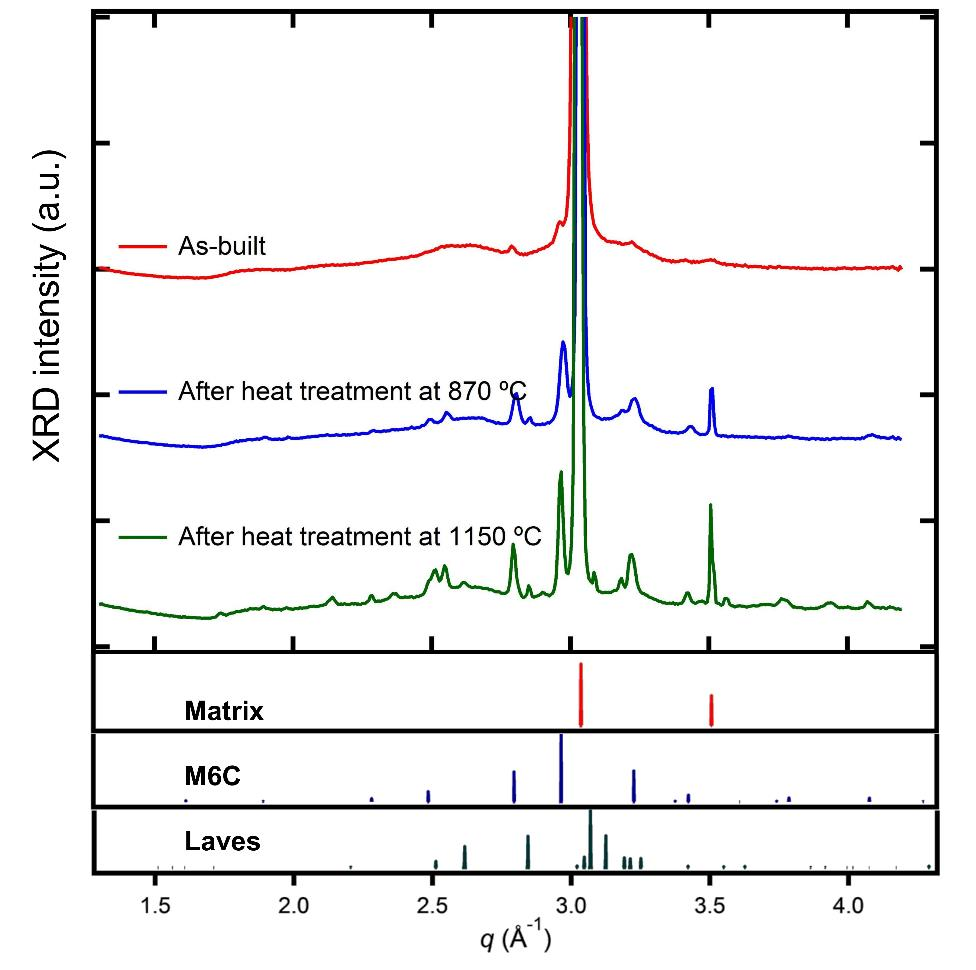}
			\caption{Synchrotron XRD observations of phase evolution during heat treatments of AM IN625
			         with scattering vector magnitude $q = 4\pi\sin(\theta)/\lambda$
			         expressed in terms of scattering angle $\theta$ and X-ray wavelength $\lambda$.
			         Top panel: experimental XRD patterns of the AM IN625 sample acquired at
			                    \SI{303}{\kelvin} following the specified heat treatments.
			         Bottom panels: simulated powder XRD patterns for dominant phases ($\gamma$, \ce{M6C}, and
			                        Laves) based on the lattice symmetry and parameters described in the text.}
			\label{fig:XRD625}
		\end{figure}

		From the room temperature XRD patterns, Figure~\ref{fig:XRD625},
		only the FCC cubic lattice ($\gamma$ phase) with a lattice parameter of
		$(0.358 \pm 0.001)$~\si{\nano\meter} could be identified.
		The first annealing step to \SI{1143}{\kelvin} led to the formation of carbides,
		most of which are \ce{M6C} with a cubic lattice of
		$Fd\bar{3}m$ group and a lattice parameter of $(1.011 \pm 0.003)$~\si{\nano\meter}.
		The second heating step to \SI{1423}{\kelvin} preserved the carbides and promoted
		the formation of an intermetallic Laves phase.
		The Laves phase has a hexagonal lattice of $P6_3/mmc$ group,
		with lattice parameters $a = (0.481 \pm 0.001)$~\si{\nano\meter}
		and $c = (1.565 \pm 0.004)$~\si{\nano\meter}.
		The uncertainties in these values are reported with \SI{95}{\percent} confidence.
		Hence, we conclude that from a statistical point of view,
		the dominant precipitates after this two-step heat treatment
		are \ce{M6C} carbides and the Laves phase.

\section{Simulation results}
	\subsection{Finite element thermal simulations\label{sec:fea_results}}
		To validate our FEA thermal model, we compared its surface temperature
		prediction against \emph{in situ} thermographic measurements.
		Details of the thermographic measurement setup were published in \cite{Lane2016},
		and data used for the single-track comparison given	here are described
		in \cite{Heigel2017}.  The same thermal camera settings used in \cite{Lane2016,Heigel2017}
		were used here: \SI{40}{\micro\second} integration time, and
		\SI{1350}{\nano\meter} to \SI{1600}{\nano\meter} spectral range. 

\begin{figure}\centering
			\includegraphics[width=\textwidth]{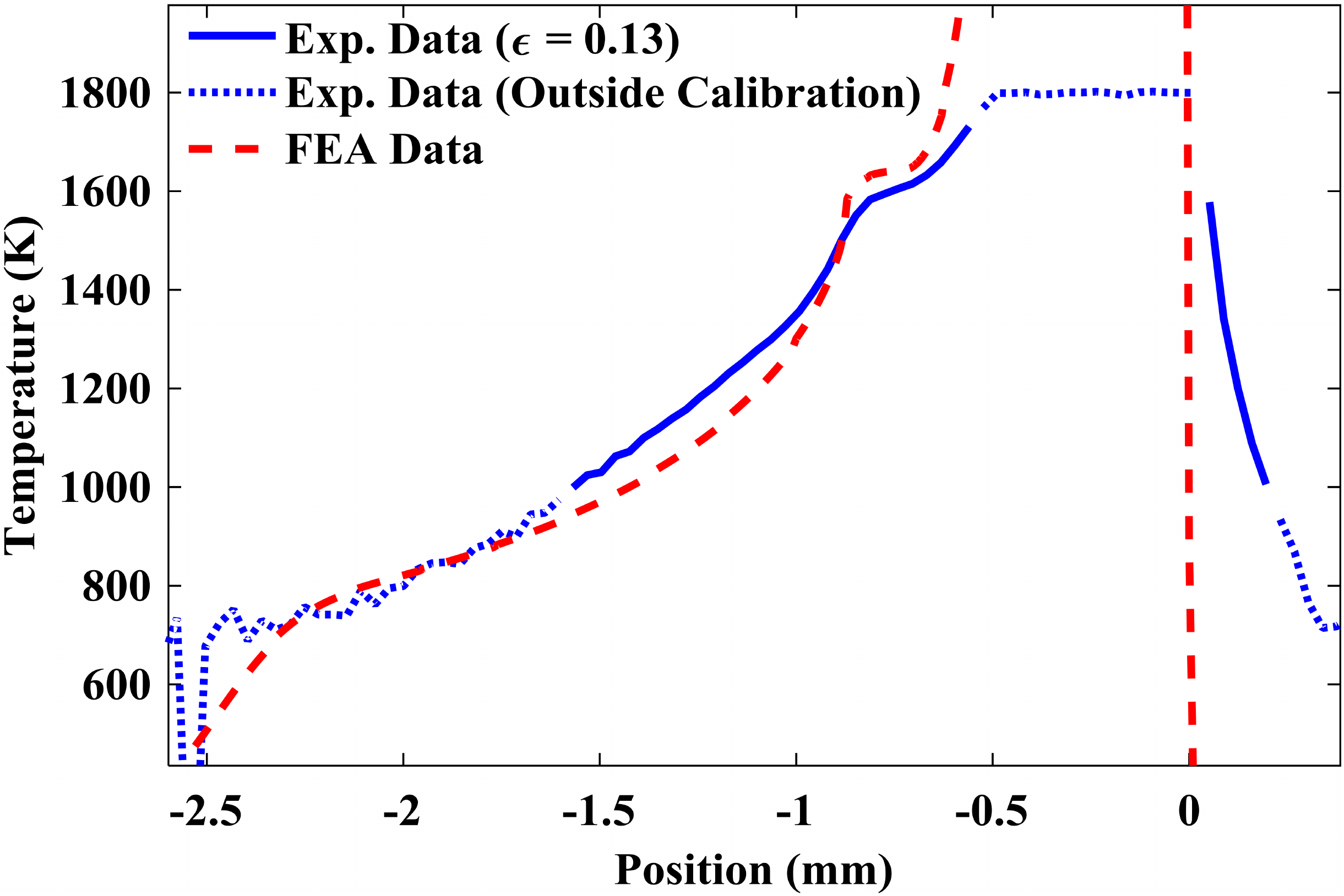}
			\caption{Results of FEA thermal modeling (red dashed line) and \emph{in situ} thermographic measurement
			         (solid and dotted blue lines) for single-line scan on bare IN625 plate.
			         Shoulder in measurement data corresponds to melt pool boundary.
			         Dotted blue lines indicate thermographic data outside the calibrated range but still shown for clarity.}
			\label{fig:heatprofile}
		\end{figure}

		As described in \cite{Lane2016}, thermographic imaging of laser scans on metal powder
		produces highly stochastic temperature fields with localized gradients due
		to the varying surface structure and emissivity, which inhibit true temperature measurement.
		In contrast, scans on flat plates of bulk metal result in smooth temperature gradients,
		and single-line scans create steady-state melt pools that simplify comparisons to FEA simulations.
		Therefore, FEA simulation results are compared against experimental laser scans
		on bare plates, without powder.
		A representative temperature profile on IN625 plate was taken from one frame of the thermal video
		after the melt pool attained nominal steady state.
		Figure~\ref{fig:heatprofile} compares the measured and predicted thermal profiles
		for a single track scan without powder using a scan speed of
		\SI{0.8}{\meter\per\second} and laser power \SI{195}{\watt}.  

		The emissivity of the surface of a real specimen---also known as emittivity---is highly	dependent
		on processing, and values for the IN625 melt pool, rolled plate, and powder surfaces are unknown.
		Therefore, the thermal camera signal cannot be directly converted to temperature
		for comparisons to FEA model predictions.
		Instead, we scale thermographic data based on an observed solidification zone following	a technique
		developed for analysis of \ce{Ti}--\SI{6}{\percent} \ce{Al}--\SI{4}{\percent} \ce{V} powder \cite{Yadroitsev2014}.
		A shoulder in the thermographic profile behind the melt pool is attributed to
		the solidification zone, apparent between \SI{-0.86}{\milli\meter} to \SI{-0.74}{\milli\meter}
		in Figure~\ref{fig:heatprofile}.
		The profile is scaled by an assumed emittivity and converted to temperature	units such that
		the temperature in this zone matches the solidus temperature, $T_s = \SI{1587}{\kelvin}$,
		which we calculated for IN625 using the TCNI8 thermodynamic database \cite{TCNI8}.
		This technique uses $T_s$ as a reference temperature, and results in
		emittivity of $\epsilon = 0.13$, which is a reasonable value for a semi-specular
		metallic surface at the same spectral range and viewing angle $(43.7^\circ)$ as
		the thermal imaging setup \cite{Touloukian1970}.
		The calibrated range of the thermal camera was \SI{823}{\kelvin} to \SI{1298}{\kelvin}.
		Temperatures above this range saturate the camera, and temperatures
		below approach the noise floor of the camera detector.
		Applying this emittivity maps the reportable temperature range to \SI{988}{\kelvin}
		to \SI{1763}{\kelvin}, as shown by the solid blue line in Figure~\ref{fig:heatprofile}.  

\begin{figure}\centering
			\includegraphics[width=\textwidth]{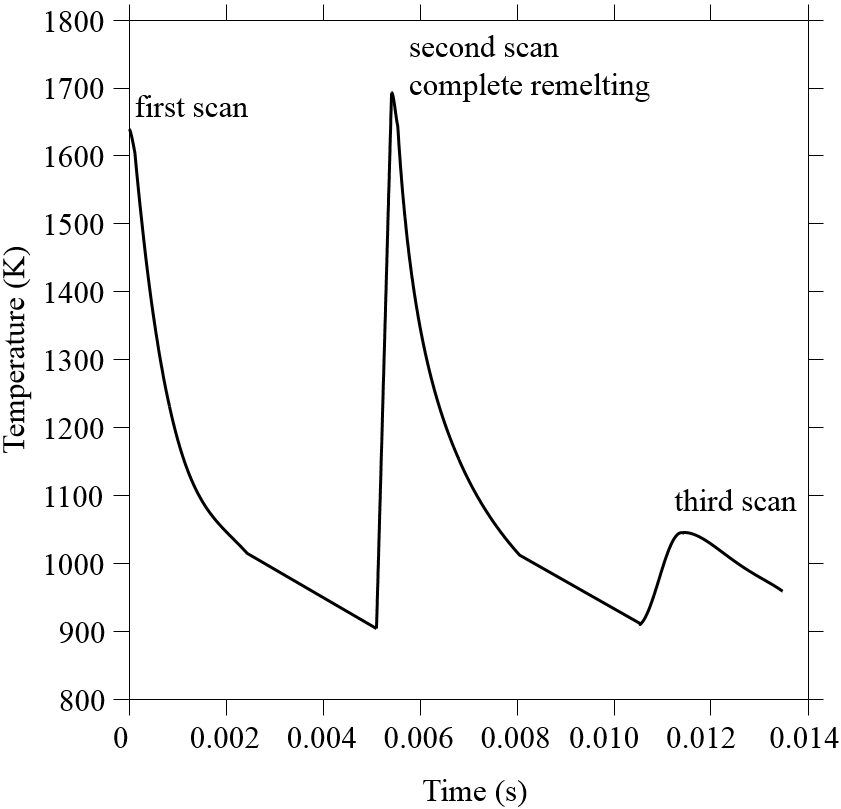}
			\caption{Temperature on the surface of a single powder layer as a function of time at
			         a position midway between two melt pool centers from a three track scanning
			         FEA thermal model. The surface material fully melts during two near passes,
			         with the molten pool overlapping this surface midpoint by \SI{20}{\micro\meter};
			         on the third pass, \SI{80}{\micro\meter} removed from the melt pool boundary,
			         this material is reheated without melting.}
			\label{fig:heatschedule}
		\end{figure}

		Figure~\ref{fig:heatschedule} shows the predicted temperature as a function of time at a point on the
		surface located midway between the centers of two of three anti-parallel laser scan tracks.
		Full melting and re-melting occurs as the laser beam traverses the two nearest tracks,
		passing the same distance away from the measurement point both times.
		Heating without melting is observed as the laser scans material along the third track.
		This three-track temperature profile was used directly in DICTRA simulations of IN625 solidification.
		
		Figure~\ref{fig:fea_meltpool} shows the thermal profile of the melt pool in cross-section.
		The rectangular meshing elements near the surface measure
		$\SI{10}{\micro\meter} \times \SI{10}{\micro\meter} \times \SI{6}{\micro\meter}$.
		Contours illustrate the temperature field, with the solid-liquid interface coinciding with
		the border between red and orange bands.
		The transformation angle $\alpha$ between the melt pool boundary and laser scanning direction
		indicates the local solidification speed, $V_s = V_b \cos\alpha$ \cite{Farzadi2008}.

		FEA thermal model results were also used to determine solidification parameters to be employed
		for the phase-field simulations described in Section~\ref{sec:pf_results}.
		Average cooling rate $\dot{T}=\SI{e5}{\kelvin\per\second}$ was calculated from
		Figure~\ref{fig:heatschedule}, measuring the slope of the line connecting the maximum and
		minimum temperatures of the second peak to its subsequent valley.
		A constant value for thermal gradient $G=\SI{e7}{\kelvin\per\meter}$ was chosen for the simulations.
		It is representative of the values along the solidus contour in Figure~\ref{fig:fea_meltpool},
		$T_s=\SI{1587}{\kelvin}$:
		we computed $G=|\nabla T|$ in the range from $\SI{0.57e7}{\kelvin\per\meter}$ to $\SI{2.2e7}{\kelvin\per\meter}$,
		for mesh points to the left of the melt pool minimum.
		Values taken along the solidus contour behind the melt pool minimum yielded a range
		of angles $\alpha$ from \SI{78}{\degree} to \SI{89}{\degree} for $V_b=\SI{0.8}{\meter\per\second}$,
		and $V_s$ ranged between \SI{0.01}{\meter\per\second} and \SI{0.17}{\meter\per\second}
		Note that the solidifying interface experiences localized thermal gradients and solidification speeds:
		the cooling rate $\dot{T} = GV_s$ in L-PBF processes is not constant.
		
	\subsection{CALPHAD-based solidification simulations}
\begin{figure*}[tp]\centering
			\includegraphics[width=\textwidth]{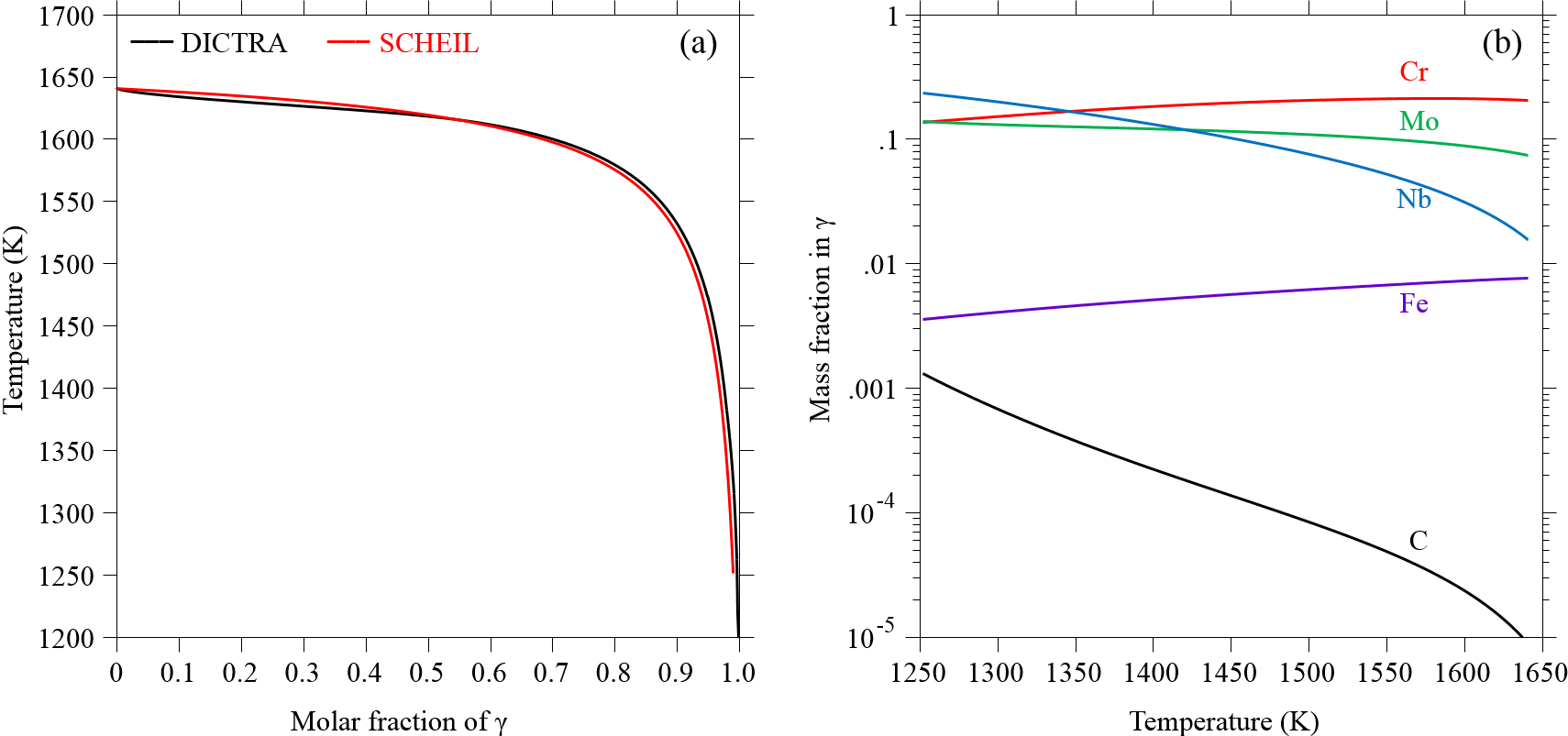}
			\caption{Effects of rapid cooling on phase transformation, using the thermal history presented
			         in Figure~\ref{fig:heatschedule}.
			         (a) Transformation of liquid into $\gamma$, expressed as molar phase fraction.
			             Low extent of solid-state diffusion under these conditions is evident in the close
			             proximity of curves from the DICTRA simulation (black line) and Scheil-Gulliver model
			             (black).
			         (b) \ce{C}, \ce{Mo}, and \ce{Nb} enrichment and \ce{Cr} depletion in $\gamma$ at
			             the growth front during solidification as predicted by the Scheil-Gulliver model,
			             demonstrating significant microsegregation.
			         Phases other than liquid or $\gamma$ are excluded from this analysis.
			        }
			\label{fig:segphase}
		\end{figure*}

		Figure~\ref{fig:segphase}a represents solidification of the interdendritic region
		as predicted by the Scheil-Gulliver model and by DICTRA for the thermal profile in
		Figure~\ref{fig:heatprofile}.
		Based on the low fraction of secondary phases observed in as-solidified experimental specimens
		(\emph{e.g.}, Figures~\ref{fig:temcarbides}~and~\ref{fig:XRD625}),
		we made the simplifying assumption that liquid solidifies into $\gamma$ phase, only.
		Due to the high cooling rate, the DICTRA results show solidification behavior very similar
		to the Scheil-Gulliver model (Figure~\ref{fig:segphase}a).
		Figure~\ref{fig:segphase}b shows the Scheil-Gulliver model prediction of liquid composition.
		Microsegregation occurs for all elements, with particularly high segregation of \ce{Mo} and \ce{Nb} into the liquid.
		In the last solidified liquid, $T=\SI{1250}{\kelvin}$, the \ce{Mo} and \ce{Nb} mass fractions are as high as
		\SI{20}{\percent} and \SI{29}{\percent}, respectively.
		The last solidified liquid is also enriched in \ce{C} whereas \ce{Cr} and \ce{Fe} are depleted.

		\begin{figure*}[tp]\centering
			\includegraphics[width=\textwidth]{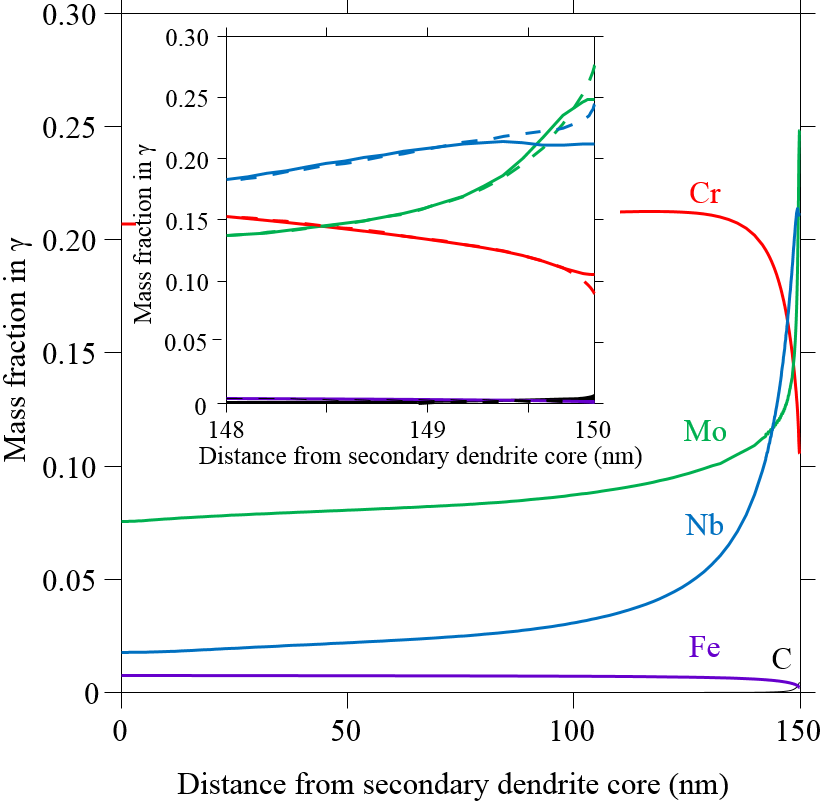}
			\caption{Composition profiles in $\gamma$ as a function of distance from the secondary dendrite core,
			         as predicted by DICTRA simulation using the thermal history presented in Figure~\ref{fig:heatschedule}.
			         Inset compares compositions after the second scan (dashed lines), during which material remelts;
			         and third scan (solid lines), which reheats without melting,
			         producing short-range diffusion (note $x$-axis scale).
			         Nearby laser scans promote solid-state diffusion, but with negligible impact on composition profiles
			         due to the short residence time at elevated temperature.
			        }
			\label{fig:segcomp}
		\end{figure*}

		Detailed DICTRA simulation results are shown in Figure~\ref{fig:segcomp},
		with an inset showing results near the centerline of interdendritic liquid.
		The composition profiles show microsegregation from the	secondary dendrite core
		($x = \SI{0}{\nano\meter}$)	to the interdendritic region ($x = \SI{150}{\nano\meter}$).
		Since applying the FEA thermal profile (Figure~\ref{fig:heatschedule}) for the DICTRA simulation
		leads to complete melting at the second scan temperature peak,
		the resulting segregation in Figure~\ref{fig:segcomp} occurs during this cooling.
		The third laser scan, centered \SI{150}{\micro\meter} from the FEA measurement point,
		leads to minor homogenization of the segregated profiles.
		This is, however, only notable over a distance less than \SI{0.5}{\nano\meter} (inset, Figure~\ref{fig:segcomp})
		and can have no significant effect on the microsegregation profile.
		Figure~\ref{fig:segcomp} shows most of the microsegregation occurs within the \SI{10}{\nano\meter}
		region near the last solidified liquid,	in good agreement with a recent synchrotron SAXS study
		of the homogenization kinetics of L-PBF IN625 \cite{Zhang2017},
		where a novel analysis directly linked the length of X-ray streaks to the length scale of the segregation.
		It can further be concluded that the segregation is in the same range as in the Scheil-Gulliver simulation.
		The main difference is that the DICTRA simulation results in a greater amount of \ce{Mo} and somewhat less
		\ce{Nb} in the interdendritic region compared to the last solidified liquid composition
		obtained from the Scheil-Gulliver model, as expected.

	\subsection{Phase-field solidification simulations}\label{sec:pf_results}
\begin{figure}[tp]\centering
		    \includegraphics[width=\textwidth]{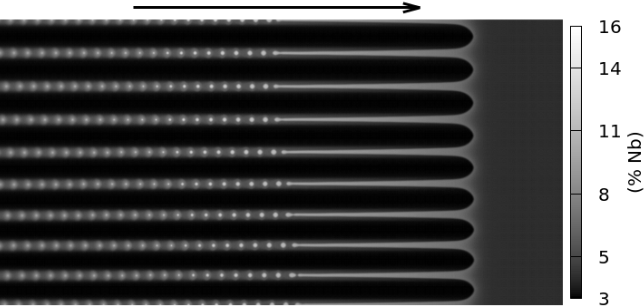}
		    \caption{Cellular microstructure in \ce{Ni}--\SI{4}{\percent}
		             \ce{Nb} predicted by phase-field
		             simulation after \SI{1.3}{\micro\second} of growth with
		             $V_s = \SI{25}{\milli\meter\per\second}$,
		             as illustrated by the scalar composition field $c$.
		             Image represents $\SI{6}{\micro\meter}\times \SI{12}{\micro\meter}$
		             window near the growth front.
		             Note steady-state \ce{Nb} enrichment of undercooled liquid near cell tips
		             and pinch-off of liquid droplets at the root of intercellular grooves.
		             \ce{Nb} diffuses out of these droplets into the surrounding solid,
		             resulting in the linear pattern of disks with fading intensity.
		             Arrow indicates growth direction.}		    	
		    \label{fig:pfresult}
		\end{figure}

		Several phase-field simulations were performed with different values for $V_s$, but constant $G$,
		in order to study microstructural evolution under various cooling conditions of interest.
		Simulations were initialized with a planar solid-liquid interface with random perturbations
		up to one grid spacing in location, and using a hyperbolic tangent profile along the growth
		axis to smooth the step change over several grid spacings for numerical stability.
		The microstructures evolved according to Equations \ref{eqn:phi}~and~\ref{eqn:conc},
		and simulations ran until steady state was achieved.
		Depending on solidification conditions, interfacial instabilities can grow into dendrites
		(with secondary arms) or cells (without).
		For the parameters chosen in this work, only cellular domains resulted,
		as represented in Figure~\ref{fig:pfresult} with $V_s = \SI{25}{\milli\meter\per\second}$.
		In Figure~\ref{fig:pfresult}, the steady-state mass fraction of \ce{Nb} in the cell center line was \SI{3.0}{\percent}.
		At the midpoint of the intercellular liquid, the mass fraction of \ce{Nb} was \SI{9.4}{\percent}.
		Intercellular liquid is shown pinching off at the root of intercellular grooves with \ce{Nb} mass fraction
		of \SI{16}{\percent}.
		This composition is below the eutectic composition of \SI{22.5}{\percent} \ce{Nb}.
		Thus secondary solid phases, which are not available to these simulations,
		are not expected to form.
		The main point is that the microsegregation found by this model is considerably less than
		is predicted by a Scheil-Gulliver analysis of binary \ce{Ni}--\ce{Nb}:
		such an analysis for $k=0.48$ predicts a mass fraction of \SI{1.9}{\percent} \ce{Nb}
		in the cell centerline, \SI{21}{\percent} in solid formed when only \SI{1}{\percent}
		liquid remains,	and likely formation of secondary phases from the last liquid to solidify.
		
\begin{figure}[tp]\centering
		    \includegraphics[width=\textwidth]{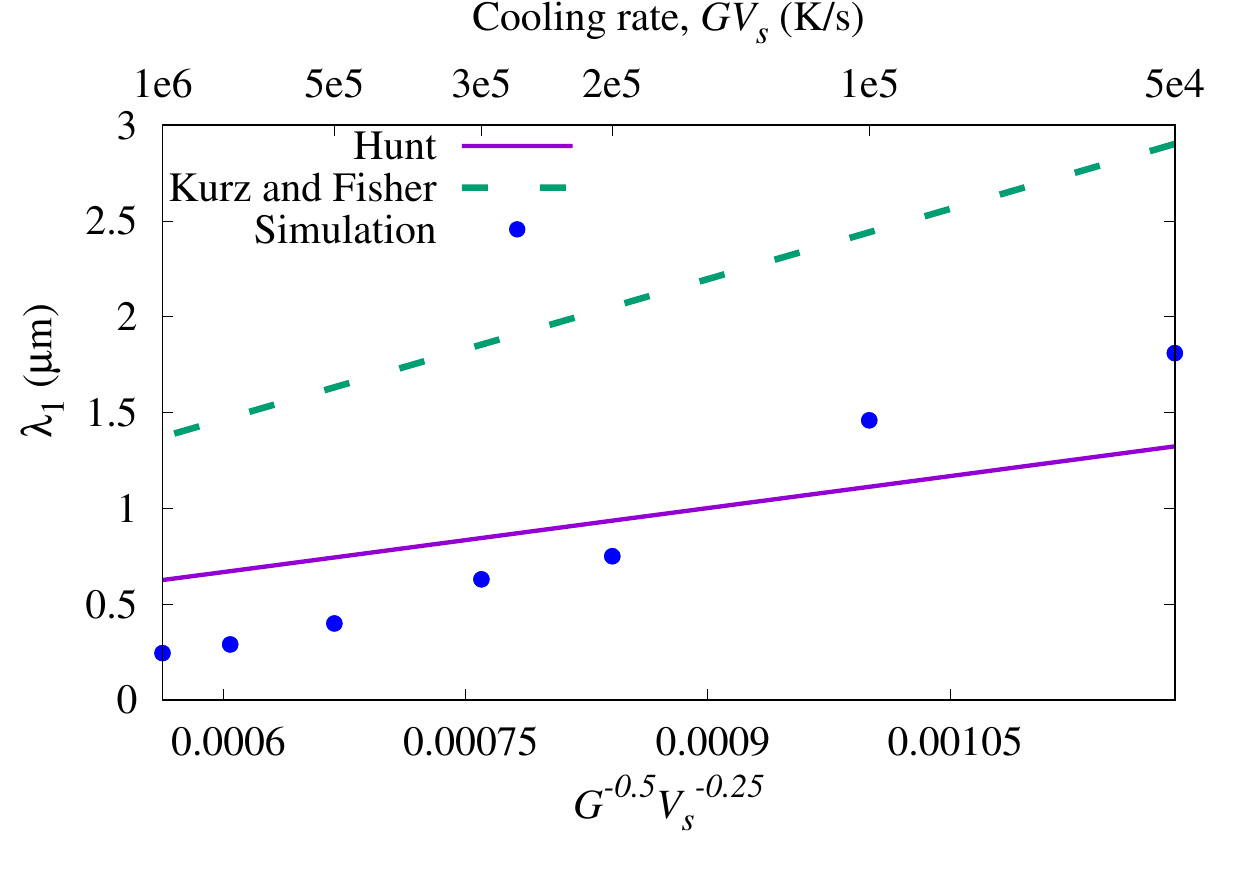}
		    \caption{Comparison of primary dendrite arm spacing measured from simulated microstructures
		             with two analytical models for simple cellular/dendritic tip geometries:
		             spherical (Hunt \cite{Hunt1979}) and ellipsoidal (Kurz and Fisher \cite{Kurz1981}).}
		    \label{fig:pdas_hunt}
		\end{figure}

		The primary dendrite arm spacing $\lambda_1$, a commonly reported metric for cellular/dendritic
		microstructures, was averaged from multiple simulations with the same $V_s$ by dividing the sum
		of simulation domain widths by the sum of the numbers of cells advancing at steady state.
		We measured $\lambda_1$ in the range from \SI{0.245}{\micro\meter} to \SI{1.81}{\micro\meter}
		from our simulations, depending on the cooling rate $\dot{T}$ which ranged from
		\SI{e4}{\kelvin\per\second} to \SI{e6}{\kelvin\per\second}, as shown in Figure~\ref{fig:pdas_hunt}.
		For reference, we compared our results with the analytical models of Hunt \cite{Hunt1979},
		\begin{equation}
			\lambda_1 =  A(k \Gamma \Delta T_0 D_\ell)^{0.25} G^{-0.5}V_s^{-0.25}
			\label{eqn:hunt}
		\end{equation}
		with $A=2.83$, and Kurz and Fisher \cite{Kurz1981},
		\begin{equation}
			\lambda_1 =  A(\Gamma \Delta T_0 D_\ell / k)^{0.25} G^{-0.5}V_s^{-0.25}
			\label{eqn:kurzapprox}
		\end{equation}
		with $A=4.3$,
		under the simplifying assumption that undercooling $\Delta T\approx\Delta T_0 = T_{\ell}-T_s$,
		measured from the equilibrium phase diagram at $c_0$.
		$\Gamma=\SI{3.65e-7}{\kelvin\meter}$ is the Gibbs-Thomson coefficient.
		The proportionality constants depend on the 3D geometry assumed for the dendrite arrays:
		Hunt assumed spherical dendrite tips and derived $A(k \Gamma \Delta T_0 D_\ell)^{0.25}$,
		while Kurz and Fisher assumed ellipsoidal tips and derived $A(\Gamma \Delta T_0 D_\ell / k)^{0.25}$.
		From the line of best fit through our simulation data points, we calculate
		$A = 6.8$ from Equation~\ref{eqn:hunt}, and	$A = 4.7$ from Equation~\ref{eqn:kurzapprox}.
		Figure~\ref{fig:pdas_hunt} shows that neither model provides an objectively good fit.
		The deviation is likely due to the combination of our reduced geometry (2D rather than 3D),
		rapid cooling $(\Delta T \neq \Delta T_0)$, and more complex tip geometry than either simple model.
		This result is supported by a body of experimental evidence demonstrating large effects
		of $c_0$, $G$, and $V_s$ on dendrite tip geometries and $\lambda_1$ \cite{Ungar1985,Liu1995,Kirkaldy1995}.

\section{Discussion}
	Experimental observations via TEM reveal significant precipitation near cell/dendrite
	boundaries in as-built L-PBF material (Figure~\ref{fig:temcarbides}).
	Stress relief at the manufacturer's recommended temperature (\SI{1143}{\kelvin}) was
	found to promote further growth of carbides, with \ce{M6C} precipitation and growth from
	the $\gamma$ matrix observed by \emph{in situ} synchrotron XRD (Figure~\ref{fig:XRD625}).
	Further treatment at \SI{1423}{\kelvin} promoted \ce{M6C} growth and Laves phase precipitation:
	after \SI{1}{\hour} at temperature, \ce{M6C} persisted and Laves phase grew to measurable levels.
	This growth is despite the fact that homogeneous IN625 of nominal composition is single-phase
	$\gamma$, as established experimentally and from thermodynamics \cite{Floreen1994,TCNI8}.

	Investigation into the cause of the observed precipitation during heat treatments
	requires knowledge of the solidification rate, which can not be directly determined from
	the laser scan speed. \emph{In situ} thermographic measurements on bare IN625
	plate provided the melt pool width and thermal gradients on the	surface.
	FEA simulations were calibrated against these surface data, then used to
	extract details of the full 3D melt pool shape during three-track L-PBF simulations
	(Figure~\ref{fig:fea_meltpool} and Section~\ref{sec:fea_results}).
	For a \ce{Yb} fiber laser operating at \SI{195}{\watt} and scanning a single powder layer
	at \SI{0.8}{\meter\per\second}, the solidification rate is between \SI{0.01}{\meter\per\second}
	at the bottom of the melt pool and \SI{0.17}{\meter\per\second} at its trailing edge.
	The cooling rate along this interface was on the order of \SI{e5}{\kelvin\per\second},
	a rapid solidification condition.

	Experimentally, these conditions produce fine cellular dendrites.
	To assess the level of microsegregation, we performed modeling on three levels:
	Scheil-Gulliver, with no diffusion in $\gamma$ and perfectly mixed liquid;
	DICTRA, with multicomponent diffusion in 1D for both $\gamma$ and liquid;
	and phase-field, with 2D cells/dendrites of $\gamma$ growing into liquid.
	
	To assess how microsegregation correlates to the experimentally identified precipitates,
	we look at the thermodynamic contribution to the driving force
	for nucleation of various phases from the $\gamma$ matrix \cite{Hillert2007},
	ignoring interfacial energy contributions and kinetic obstacles.
	The phase with largest driving force would be expected to nucleate first.
	Comparison of these driving force values may provide additional insight
	regarding the nucleation of precipitates in dendritically segregated material.
	In Table~\ref{tab:drivingforce}, the phases most likely to nucleate are
	listed in decreasing order of driving force for the composition of solid when \SI{1}{\percent} liquid remains
	(mass fraction \ce{Ni}--\SI{0.13}{\percent} C--\SI{13.6}{\percent} Cr--\SI{0.35}{\percent} \ce{Fe}--\SI{13.9}{\percent} \ce{Mo}--\SI{23.5}{\percent} \ce{Nb}).
    On this basis, we expect \ce{Nb}-rich \ce{MC} carbide to nucleate first;
    if it cannot, we expect \ce{M2C}.
    If \ce{M2C} cannot form, a \ce{Mo}-rich phase is expected, \ce{M6C} or BCC,
    followed by \ce{Cr}-rich $\sigma$, $P$, Laves, or \ce{M23C6}.
    As Table~\ref{tab:drivingforce} demonstrates, the specific nucleation order depends on temperature.
    Note that this is a sequence of energetic favorability, not a predicted time-evolution:
    phases toward the bottom of the table can only precipitate if local conditions preclude
    any of the more-favorable phases from doing so.
    The favored precipitate is also not the same everywhere: local variations in composition
    and other materials properties will change the tabulated sequence.
    Competitive growth and coarsening may occur when time and diffusion are factored in,
    but these effects are beyond the scope of this paper.

	\begin{table}\centering
		\caption{Thermodynamic driving force for nucleation of secondary phases from $\gamma$
		         for the enriched (interdendritic) composition,
		         \ce{Ni}--\SI{0.13}{\percent} C--\SI{13.6}{\percent} Cr--\SI{0.35}{\percent} \ce{Fe}--\SI{13.9}{\percent} \ce{Mo}--\SI{23.5}{\percent} \ce{Nb},
		         at the stress relief and homogenization treatment temperatures.
		         Greater values indicate larger driving forces for nucleation.
		         Equilibrium phases for each temperature are underlined.}
		\label{tab:drivingforce}

		\begin{tabular}{|lr|lr|}\hline
			\multicolumn{2}{|c|}{\SI{1143}{\kelvin}} & \multicolumn{2}{c|}{\SI{1423}{\kelvin}} \\
			Phase & \multicolumn{1}{c|}{$-\Delta G^{\mathrm{nuc}}$} & Phase & \multicolumn{1}{c|}{$-\Delta G^{\mathrm{nuc}}$}  \\\hline
			$\eqphase{MC}$     & \SI{20.5}{\kilo\joule\per\mole}  & $\eqphase{MC}$     & \SI{17.0}{\kilo\joule\per\mole}\\
			\ce{M2C}           & \SI{15.6}{\kilo\joule\per\mole}  & \ce{M2C}           & \SI{13.1}{\kilo\joule\per\mole}\\
			$\eqphase{\mu}$    & \SI{8.0}{\kilo\joule\per\mole}   & \ce{M6C}           & \SI{5.8}{\kilo\joule\per\mole} \\
			\ce{M6C}           & \SI{7.9}{\kilo\joule\per\mole}   & BCC                & \SI{3.3}{\kilo\joule\per\mole} \\
			BCC                & \SI{6.3}{\kilo\joule\per\mole}   & $\eqphase{\mu}$    & \SI{3.0}{\kilo\joule\per\mole} \\
			$\eqphase{\sigma}$ & \SI{5.2}{\kilo\joule\per\mole}   & $\eqphase{Laves}$  & \SI{2.9}{\kilo\joule\per\mole} \\
			Laves              & \SI{4.1}{\kilo\joule\per\mole}   & liquid             & \SI{1.8}{\kilo\joule\per\mole} \\
			$\eqphase{\delta}$ & \SI{3.5}{\kilo\joule\per\mole}   & $\eqphase{\sigma}$ & \SI{1.2}{\kilo\joule\per\mole} \\
			$\gamma''$         & \SI{3.5}{\kilo\joule\per\mole}   & $\gamma''$         & \SI{1.2}{\kilo\joule\per\mole} \\
			\ce{M23C6}         & \SI{3.4}{\kilo\joule\per\mole}   & $\eqphase{\delta}$ & \SI{1.1}{\kilo\joule\per\mole} \\\hline
		\end{tabular}
	\end{table}
	
	The nucleation sequence suggested by the driving force calculations can be compared to the
	calculated equilibrium phases for these compositions at the stress relieving and homogenizing temperatures.
	Thermodynamic calculations show \ce{MC}, BCC, $\delta$ and $\sigma$ phases to be stable at both temperatures,
	while $\mu$ phase is only stable at the lower temperature.
	Therefore, as the highly segregated profiles diffuse during long heat treatments,
	solid state transformations are to be expected---an active topic of research beyond the scope of this paper.
	
	It is important to note that the nucleation sequence and the ability for a phase
	to form are dependent on several mechanisms and not only the nucleation
	driving force. A complete description of nucleation must include
	the interfacial energy between the matrix and secondary phase, which depends
	on the interfacial area and crystallographic details.
	For large or complex unit cells in particular, kinetic effects play an important role:
	a thermodynamically stable phase may be unable to nucleate or grow if its constituent elements
	are not sufficiently mobile in the matrix at the specified temperature and composition.
	This may explain, for example, the fact that BCC phase has not been observed in our
	experiments despite its apparently high driving force for nucleation.
	Furthermore, the thermodynamic database used for the driving force calculation
	may contain uncertainties and is also limited to a certain energy resolution. 
	This could, in particular, be the case when different phases have
	similar driving forces, e.g. \ce{M6C} and BCC in our case.

	Even if \ce{M6C} and Laves are not equilibrium phases at these conditions,
	their driving forces for nucleation	are high enough to put them early in the nucleation sequence:
	earlier than the equilibrium phases $\delta$ and $\sigma$.
	This could explain the experimental observation of these phases.
	It is also interesting to note that the \ce{Nb}-rich \ce{MC} carbide is an 
	equilibrium phase at temperatures as high as \SI{1423}{\kelvin},
	which could complicate the homogenization treatment.

	The binary phase-field simulations clearly show microsegregation in the cellular growth pattern.
	The spherical droplets in Figure~\ref{fig:pfresult} form at high solidification speeds
	as the advancing cells pinch off liquid pockets to maintain	the steady-state intercellular groove depth.
	These liquid droplets are enriched to more than twice the initial \ce{Nb} composition,
	depending on cooling rate, with some loss to diffusion during solidification.
	Experiments involving rapid solidification of electron-beam melted \ce{Al}--\ce{Fe}--\ce{Ni} alloys produced
	morphologically similar droplets, which precipitated intermetallic compounds during solidification \cite{Boettinger1988}.
	While our models did not consider secondary solid phases,
	and therefore could not simulate precipitation in the evolving microstructure,
	the microsegregation patterns are real and significant.
	It is interesting to note that even the highly concentrated droplets are enriched less than
	half as much as the microsegregation predicted by the DICTRA simulations.
	To some extent, this difference reflects model geometry:
	the phase-field simulations were performed in 2D with realistically curved dendrite tips,
	while DICTRA simulations represent 1D planar solidification.
	Curvature effects significantly affect diffusion.
	In addition, under rapid solidification conditions, local chemical equilibrium across the solid-liquid interface
	is not expected to hold, and results in the natural phenomenon known as solute trapping \cite{Baker1969}.
	DICTRA enforces local equilibrium without accounting for solute trapping,
	and therefore over-estimates the extent of microsegregation in these simulations.
	While the phase-field model does not enforce local equilibrium, the diffuse interface artificially magnifies the effect.
	Our phase-field model includes an anti-trapping flux in Equation~\ref{eqn:conc}, to correct the spurious contribution \cite{Karma1998}.
	The correction is accurate to second-order in $\frac{W_0}{d_0}$,
	the dimensionless ratio of interface thickness to capillary length,
	and is exact for low solidification velocities.
	Under the rapid solidification conditions investigated here, this second-order expression
	does not cancel the spurious effect, which artificially decreases the simulated microsegregation.
	Under L-PBF conditions, which produce both curved interfaces and solute trapping in real material,
	the phase-field model produces a more accurate estimate	of microsegregation than DICTRA for the the cellular/dendritic microstructure.

	There are also confounding factors that affect the exact compositions predicted
	by our phase-field and DICTRA models.
	The real temperature-time cycle during L-PBF is more complex than the FEA model
	employed in this study: for example, the effects of building additional layers
	on the temperature profile are neglected here.
	This simplified thermal history may discount the influence of solid state diffusion,
	producing artificially high compositions in enriched regions of the microstructure.
	This work also models the laser input as a surface heat source, only, which neglects
	the effects of particle ejection on heat and mass transfer in the melt pool \cite{Khairallah2016,Teng2016}. 
	Furthermore, thermophysical properties such as $k$ and $m_{\ell}$ depend on the
	solidification speed, and may deviate significantly from equilibrium values in the
	rapid solidification regime, a topic explored in greater detail elsewhere \cite{Ghosh2017}.
	Finally, powder grains of somewhat different sizes and composition lead to 
	compositional variations throughout the part, which may produce highly
	localized segregation patterns not captured in our models.

\section{Conclusion}
	From this work, we conclude:
	\begin{itemize}
		\item For additively manufactured Inconel 625, heat treatments of
		      \SI{1}{\hour} at \SI{1143}{\kelvin} and \SI{1423}{\kelvin}
		      promote precipitation of secondary phases from the $\gamma$ matrix.
		\item FEA simulations of the moving 3D melt pool shape predict maximum
		      solidification rates only \num{0.2} times the laser scan speed,
		      accompanied by a cooling rate of \SI{e5}{\kelvin\per\second}.
		      This L-PBF process occurs under rapid solidification conditions.
		\item The Scheil-Gulliver model was used to compute the maximum possible
		      extent of microsegregation for the multicomponent alloy IN625.
		      DICTRA simulations that include the role of diffusion in $\gamma$
		      reduce the degree of microsegregation near the end of solidification.
		      Phase-field simulations of cellular/dendritic microstructures further
		      reduce microsegregation, finding \ce{Nb} compositions \num{0.4} times
		      the Scheil-Gulliver prediction.
		\item Carbides (\ce{MC}, \ce{M2C}, and \ce{M23C6}), topologically close-packed
		      ($\delta$, $\mu$, $\sigma$), and Laves phases have negative thermodynamic
		      driving forces for nucleation from the FCC phase under the conditions of
		      microsegregation and temperature studied: these may spontaneously precipitate
		      under the conditions of temperature and microsegregation investigated.
	\end{itemize}

	This effort demonstrates the viability of cooperatively modeling L-PBF processes
	using several techniques to assess microstructural phenomena in IN625.
	Work in progress will extend this tool chain to investigate the effects of interfacial energy
	and diffusion in microsegregated regions on competitive precipitation
	and solid state transformations.
	With a complete microstructure model, we plan to evaluate the whole process
	in order to find a better way to prepare this material for service.

\section*{Acknowledgment}
Use of the Advanced Photon Source, an Office of Science User Facility operated
for the U.S. Department of Energy (DOE) Office of Science by
Argonne National Laboratory, was supported by the U.S. DOE under
Contract No.~DE-AC02-06CH11357.

\bibliographystyle{elsarticle-num}
\bibliography{Acta_Additive625}

\end{document}